\documentclass[conference]{IEEEtran}

\usepackage{graphicx}
\usepackage{times}
\usepackage{subfigure}
\usepackage{amsfonts}
\usepackage{amsmath,amssymb}
\usepackage{ams}

\DeclareGraphicsExtensions{.jpg, .eps}

\begin{document}

\title{Reconciling Synthesis and Decomposition:\\ A Composite Approach to Capability Identification}


\author{
\authorblockN{Ramya Ravichandar and James D. Arthur}
\authorblockA{Department of Computer Science\\
Virginia Polytechnic Institute and State University\\
Blacksburg, Virginia 24060\\
Email: \{ramyar, arthur\}@vt.edu}
\and
\authorblockN{Robert P. Broadwater}
\authorblockA{Electrical and Computer Engineering\\
Virginia Polytechnic Institute and State University\\
Blacksburg, Virginia 24060\\
Email: \{dew\}@vt.edu}
}

\maketitle
\begin{abstract}
Stakeholders' expectations and technology constantly evolve during the lengthy development cycles of a large-scale computer based system. Consequently, the traditional approach of baselining requirements results in an unsatisfactory system because it is ill-equipped to accommodate such change. In contrast, systems constructed on the basis of Capabilities are more change-tolerant; Capabilities are functional abstractions that are neither as amorphous as user needs nor as rigid as system requirements. Alternatively, Capabilities are aggregates that capture desired functionality from the users' needs, and are designed to exhibit desirable software engineering characteristics of high cohesion, low coupling and optimum abstraction levels. To formulate these functional abstractions we develop and investigate two algorithms for Capability identification: Synthesis and Decomposition. The synthesis algorithm aggregates detailed rudimentary elements of the system to form Capabilities. In contrast, the decomposition algorithm determines Capabilities by recursively partitioning the overall mission of the system into more detailed entities. Empirical analysis on a small computer based library system reveals that neither approach is sufficient by itself. However, a composite algorithm based on a complementary approach reconciling the two polar perspectives results in a more feasible set of Capabilities. In particular, the composite algorithm formulates Capabilities using the cohesion and coupling measures as defined by the decomposition algorithm and the abstraction level as determined by the synthesis algorithm.
\end{abstract}

\section{Introduction}
\label{sec:Introduction}
The property of change-tolerance is of paramount importance in complex emergent systems. These computer based systems are of large magnitude, have lengthy development cycles and are envisioned to be utilized for an extended lifetime. In addition, their inherent complexity results in  emergent behavior \cite{Heylighen1989} that is often unexpected. For example, the introduction of a new functionality in the system may result in unanticipated interactions with other existing components that can be detrimental to the overall system functionality. Moreover, in order to function satisfactorily complex emergent systems must accommodate the effect of dynamic factors such as varying expectations of the stakeholders, changing user needs, technology advancements, scheduling constraints and market demands, during their lengthy development periods. We conjecture that these changes can be accommodated with minimum impact, if systems are architected using aggregates that are embedded with change-tolerant characteristics. We term such aggregates as \textit{Capabilities}. Capabilities are functional abstractions that exhibit high cohesion, low coupling and balanced abstraction levels. The property of high cohesion helps localize the impact of change to within a Capability. Also, the ripple effect of change is less likely to propagate beyond the affected Capability because of its reduced coupling with neighboring Capabilities. An optimum level of abstraction assists in the understanding of the functionality in terms of its most relevant details \cite{Parnas1984}. In addition, we observe that the abstraction level is related to the size of a Capability; the higher the abstraction level, the greater is the size of a Capability \cite{Ravichandar2007}. From a software engineering perspective, abstractions with a smaller size are more desirable for implementation. Therefore, we need to design an algorithm based on the three characteristics of cohesion, coupling and abstraction, that in some sense, ``optimizes'' the identification of Capabilities. Specifically, we use a top-down and a bottom-up approach as the basis of the algorithms for formulating  Capabilities. This is because our cognitive ability to examine a problem from both a top-down and a bottom-up perspective facilitates the application of widely diverse solution approaches. This phenomenon is evident in the field of software engineering where development strategies such as top-down design, bottom-up testing, top-down integration and others that incorporate a top-down or a bottom-up perspective are utilized in the different stages of system development. In particular, for Capability identification we focus on needs analysis, a phase prior to requirements specification, because Capabilities are formulated from user needs. At this point we consider only the functional aspects of the system. Following convention, we develop two algorithms for Capability identification that are based on the top-down and bottom-up approaches:
\begin{itemize}
	\item \textit{Synthesis:} This is an algorithm based on the bottom-up approach. The system is understood in terms of its most detailed elements, which are then systematically aggregated to form   abstractions of higher levels. 
	\item \textit{Decomposition:} This is an algorithm based on the top-down approach. The system is visualized in terms of its highest level mission, which is then systematically decomposed into abstractions that are more detailed. 
\end{itemize}
In either approach the objective is to identify functional abstractions that are maximally cohesive and minimally coupled as Capabilities. We assessed the efficacy of the synthesis and decomposition algorithms by executing them on a real-world computer based library system. Our empirical analysis reveals that neither approach is sufficient by itself to determine the best set of Capabilities. More specifically, the cohesion measure based on the synthesis approach is inordinately subjective. Additionally, the synthesis strategy provides little information to assist coupling measurements. However, this approach identifies aggregates of reduced sizes as Capabilities. In contrast, the decomposition approach expedites the measurement of cohesion and coupling but results in Capabilities that are of increased sizes. In other words, these Capabilities are defined at very high levels of abstraction. Therefore, we construct a composite algorithm to establish an equilibrium between the two polar approaches. This algorithm is based on a complementary approach that incorporates elements of cohesion and coupling from the decomposition strategy, and models abstraction from the synthesis perspective. 

The remainder of the paper is organized as follows: Section \ref{sec:CapabilitiesEngineering} discusses related work and outlines the overall process of engineering Capabilities. In Section \ref{sec:Synthesis} and Section \ref{sec:Decomposition}, we compare and contrast the three primary elements that determine a Capability --- cohesion, coupling, and abstraction level --- from the synthesis and decomposition approaches, respectively. In Section \ref{sec:Reconciliation}, we describe our composite algorithm that combines the two approaches. Our conclusions  are presented in Section \ref{sec:Conclusion}.

\section{Background}
\label{sec:CapabilitiesEngineering}
A system operating in the real world is subject to dynamic factors of change. These factors necessitate system evolution, the process of constantly adapting to various influences in order to function satisfactorily \cite{Lehman1996}. Software development processes that are ill-equipped to accommodate change are primarily afflicted with requirements volatility \cite{Harker1993}. This phenomenon is known to increase the defect density and affect project performance resulting in schedule and cost overruns \cite{Malaiya1999} \cite{Zowghi2002}. Traditional Requirements Engineering (RE) strives to manage volatility by baselining requirements. However, the dynamics of user needs and technology advancements during the extended development periods of complex emergent systems discourage fixed requirements. More recently, techniques such as the Performance based specifications \cite{Performance1999} \cite{Performance2000} and Capability Based Acquisition (CBA) 
\cite{Montroll2003} are being utilized to mitigate change in large-scale systems. Performance based specifications are requirements describing the outcome expected of a system from a high-level perspective. The less-detailed nature of these specifications provides latitude for incorporating appropriate design techniques and new technologies. Similarly, CBA is expected to accommodate change and produce systems with relevant capability and current technology. It does so by delaying requirement specifications in the software development cycle, and by allowing time for a promising technology to mature so that it can be integrated into the software system.
However, the Performance based specification and the CBA approaches lack a scientific procedure for deriving system specifications from an initial set of user needs. Moreover, they neglect to define the level of abstraction at which a specification or Capability is to be described. 
Thus, these approaches propose solutions that are neither definitive, comprehensive nor mature enough to accommodate change and benefit the development process for complex emergent systems.

\begin{figure}[htbp]
\centering
\includegraphics[trim = 8mm 180mm 0mm 13mm, clip, width=10 cm]{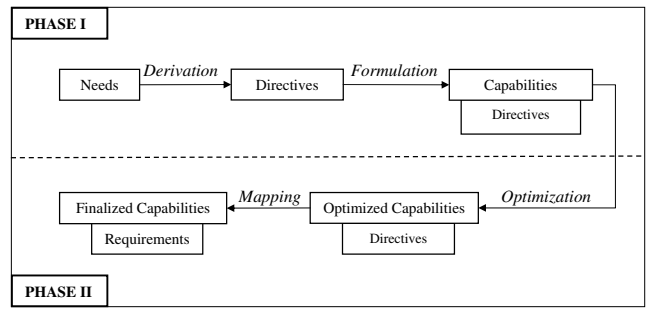}
\caption{\em Capabilities Engineering Process}
\label{fig:FigNeedsToRequirements}
\end{figure}

Our approach, the Capabilities Engineering (CE) process, architects change-tolerant systems on the basis of optimal sets of Capabilities. In fact, Rowe and Leany suggest that it is beneficial to address the issues of evolution when modeling the system architecture \cite{Rowe1997}. Therefore, we design Capabilities to incorporate evolutionary-friendly characteristics such as high cohesion, minimal coupling, and pragmatic levels of functional abstraction. Figure \ref{fig:FigNeedsToRequirements} illustrates the two major phases of the CE process.
Phase I identifies sets of Capabilities based on the values of cohesion, coupling and abstraction levels.
Techniques of modularization suggest that high cohesion and low coupling are typical of stable units \cite{Parnas1972} \cite{Yourdon1979}. Stability implies resistance to change; in the context of CE, we interpret stability as a property that accommodates change with minimum ripple effect. Ripple effect is the phenomenon of propagation of change from the affected source to its dependent constituents.
Specifically, dependency links between aggregates behave as change propagation paths. The higher the number of links, the greater is the likelihood of ripple effect. Because coupling is a measure of interdependence between units \cite{Stevens1974} we choose coupling as one indicator of stability of an aggregate. In contrast, cohesion --- the other characteristic of a stable structure --- depicts the ``togetherness'' of elements within an aggregate. A unit is said to be highly cohesive if each of its elements is directed towards achieving a single objective. As a general observation as the cohesion of a unit increases, the coupling between the units decreases. However, this correlation is only approximate, and thereby, cannot be used to estimate the values of cohesion and coupling \cite{Yourdon1979}. Therefore, we develop specific metrics to compute these values for potential Capabilities. 

Phase II, a part of our ongoing research, further optimizes these initial sets of Capabilities to accommodate schedule constraints and technology advancements. In this paper, we focus on identifying Capabilities as outlined by Phase I.

In the following sections, we discuss the synthesis and the decomposition algorithms for computing Capabilities. We then explain the necessity for a composite algorithm that includes elements of cohesion, coupling, and abstraction from both these approaches.

\section{Synthesis}
\label{sec:Synthesis}

The objective of the synthesis algorithm is to formulate Capabilities --- functional abstractions with high cohesion and low coupling --- from user needs that are obtained during the process of elicitation \cite{Goguen1993}. Needs are affiliated with the problem domain and requirements are associated with the solution domain. Capabilities are computed after the analysis of user needs but prior to requirements specification. We envision that by doing so Capabilities can bridge the chasm  between the problem and the solution space, also described as the complexity gap \cite{Racoon1995}. It is recognized that this gap is responsible for information loss, misconstrued needs, and other detrimental effects that plague system development \cite{Zave1996} \cite{Vinter2001}.
The synthesis algorithm is based on a bottom-up approach, and hence, envisions a system in terms of its details. In particular, we consider system details that are defined at low levels of abstraction and are stated from a user's perspective. We term these details as \textit{directives}. More specifically, a directive is a system specification that is described using the terminology of the problem domain. In contrast, a requirement is a system specification stated in the technical language of the solution domain. However, both a directive and a requirement share the commonality of being defined at a low level of abstraction.

Directives are a natural derivative of user needs. We use the directives as input to the synthesis algorithm for formulating Capabilities because they serve three main purposes. 
Firstly, directives strive to alleviate loss of domain knowledge, which has been identified as an important problem in RE \cite{Zave1996}. They do so by describing system functionality in terms of the problem domain. This assists in capturing domain information.
Secondly, directives are utilized to compute the cohesion and coupling values of potential Capabilities. Recall that optimal sets of Capabilities are to be determined from different functional abstractions.
Capabilities are essentially system functionalities, and hence, are associated with one or more directives. Therefore, the cohesion and coupling measures of Capabilities are determined using directives. 
Lastly, directives facilitate the mapping to system requirements. Note that Capabilities only provide a high-level architecture based on system functionalities, and therefore, requirement specifications are still necessary to direct system development. Thus, directives are easily mapped to requirements because both entities are defined at similar levels of abstraction.

\subsection{Algorithm}
\label{sec:SynAlgorithm}

\begin{figure}[htbp]
\centering
\includegraphics[trim = 0mm 162mm 0mm 10mm, clip, width=9 cm]{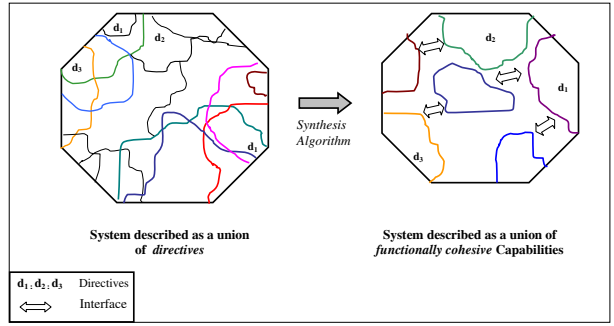}
\caption{\em Objective of Synthesis Algorithm}
\label{fig:Fig-Objective-Synthesis-Algorithm}
\end{figure}

The synthesis algorithm aims to identify abstractions with maximum cohesion and minimum coupling, as Capabilities. In particular, it strives to maximize functional cohesion, the most desirable cohesion among all other types of cohesion (coincidental, logical, temporal, procedural, communicational, and sequential) \cite{Bieman1994}. This objective of the synthesis algorithm is illustrated in Figure \ref{fig:Fig-Objective-Synthesis-Algorithm}. 
If every element of a unit is essential to the performance of a single function, then that unit is said to exhibit high functional cohesion \cite{Yourdon1979}. Therefore, the first step of the algorithm enumerates functions that possess high functional cohesion. More specifically, we examine the significance of  each directive in accomplishing various system functions. We use these significance values to compute the cohesion of a function in terms of all its participating directives.
However, it is possible that the same function is described at different levels of abstraction. We represent the functions using Venn diagrams to visually understand and resolve the discrepancies in the abstraction levels. The algorithm is explained in detail next.

\begin{figure}[htbp]
\centering
\includegraphics[trim = 20mm 218mm 0mm 25mm, clip, width=10 cm]{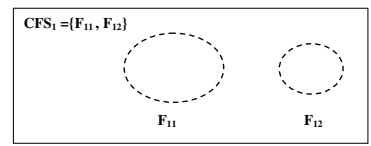}
\caption{\em Example CFS for directive $d_1$}
\label{fig:Fig-CoreFunctionSet}
\end{figure}

Let $d_1,d_2,\ldots,d_n$, $n \in \N$, denoting directives derived from user needs be the input to the synthesis algorithm. For each $d_i$ perform the following steps to determine the Capabilities of a system:

\begin{enumerate}

\item \textit{Identify all possible functions to which directive $d_i$ contributes.} 
The relevance of a directive in accomplishing a function is estimated using the impact categories shown in Table \ref{tab:ScaleOfRelevance}. This classification is intended to assess the impact of risks on a project \cite{Boehm1989}. The failure to implement a directive is also a risk, and thereby, we use this classification to determine the significance of a directive in implementing a system functionality. We assign relevance values based on the perceived significance of each impact category; these values are normalized to the [0,1] scale. 

\begin{table}[htbp]
\centering
\caption{\em Relevance Values}
\begin{tabular}{|l|l|c|}
\hline
\textsc{Impact} & \textsc{Description} & \textsc{Relevance}  \\
\hline \hline

Catastrophic &  Task failure & 1.00\\
\hline

Critical & Task success questionable & 0.70\\
\hline

Marginal &  Reduction in performance & 0.30\\
\hline

Negligible& Non-operational impact  & 0.10\\
\hline
\end{tabular}
\label{tab:ScaleOfRelevance}
\end{table}
Formally, we enumerate the list of functions $f_{im}$, $m<n$, that $d_i$ is associated with, as $InitialSet_i = \{f_{i1},f_{i2},$ $\ldots,f_{im}\}$. For example, let $d_1$ help achieve functions 
$f_{1j}, j=1, \ldots, 7$. A Venn diagram representation indicating the different abstraction levels of the functions of $InitialSet_1$ is shown in Figure \ref{fig:Fig-Directive-Snapshot}. Late, we use the relevance values of directives later to compute the cohesion of potential Capabilities.
\begin{figure}[htbp]
\centering
\includegraphics[trim = 0mm 181mm 10mm 10mm, clip, width=9 cm]{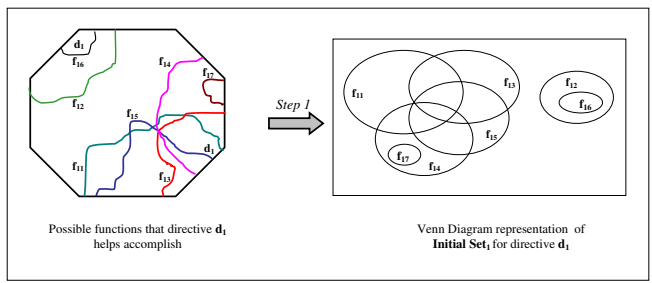}
\caption{\em Example Initial Set for directive $d_1$}
\label{fig:Fig-Directive-Snapshot}
\end{figure}

\item \textit{Expected system functionalities deduced from user needs can be stated at different levels of abstraction.} Consequently, certain functions constituting $InitialSet_i$ may be inclusive of other functions in the same $InitialSet_i$. For example, in Figure \ref{fig:Fig-Directive-Snapshot}, $f_{12}$ is inclusive of $f_{16}$. 
We avoid considering functional abstractions that are partially or completely redundant as potential Capabilities by constructing $Subset_i \subseteq InitialSet_i$ where			
$Subset_i=\{f_{ix}|f_{ix} \supseteq f_{iy}, \forall f_{iy} \in InitialSet_i; 1 \leq x,y \leq m\}$.
Note that the functions in $Subset_i$ are not encompassed by any other function in $InitialSet_i$. This implies that $Subset_i$ consists of	functions defined at the \textit{highest} level of abstraction among all other functions in $InitialSet_i$. Thus, as shown in Figure \ref{fig:Fig-Synthesis-Subset} for $d_1$, $Subset_1=\{f_{1j}\}, j=1, \ldots, 5$.

\begin{figure}[htbp]
\centering
\includegraphics[trim = 20mm 200mm 20mm 22mm, clip, width=10 cm]{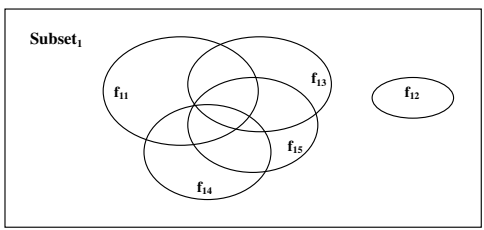}
\caption{\em Example Subset for directive $d_1$}
\label{fig:Fig-Synthesis-Subset}
\end{figure}

\item \textit{Although the aggregates in $Subset_i$ are not subset to any other aggregate, they can share common functionalities, which is an indicator of coupling.} Recall that a Capability is a self-contained functional abstraction that is minimally coupled with other Capabilities. We strive to minimize the coupling between abstractions by reducing their dependencies. Specifically, in the synthesis algorithm we use the abstraction level as an instructive factor in constructing minimally coupled aggregates. The technique of abstraction allows us to contain the dependencies within the boundaries of a higher abstraction. In particular, we identify aggregates that exhibit overlapping functionalities and aggregate them to form more decoupled abstractions. Hence, we create \textit{aggregate subsets} $AG_{ij} (1\leq i \leq n; 1\leq j \leq m)$ from $Subset_i$ to contain aggregates with commonalities. Specifically,
\begin{center}
 $AG_{ij} = \{f_{ix}, f_{iy} | f_{ix} \cap f_{iy} \neq \emptyset;1 \leq x,y \leq m\}$
\end{center}
 such that $Subset_i =  \underset{\scriptstyle x} \cup f_{ix}, \forall f_{ix}\in AG_{ij}$;
 
We then abstract the entities of $AG_{ij}$ to form higher level aggregates such that
$AG_{ij} = \{F_{ij} \}$ where $F_{ij} = \{f_{ix} \cup f_{iy} \cup \ldots \cup f_{iz}\}; 1 \leq x,y, \ldots, z \leq m$.
$F_{ij}$ encompasses all aggregates in $AG_{ij}.$ We term $F_{ij}$ as \textbf{core functions}. Hence, we utilize core functions to derive and represent the functionality of system aggregates at a higher level of abstraction. For example, for directive $d_1$, in Figure \ref{fig:Fig-AggregateSubset}, $AG_{11}=\{F_{11}\}$ where $F_{11}=\{\underset{\scriptstyle j} \cup f_{1j}\}, j=1, 3, 4, 5$ and 
$AG_{12}=\{F_{12}\}$ where $F_{12}=\{f_{12}\}$.
		
\begin{figure}[htbp]
\centering
\includegraphics[trim = 15mm 213mm 30mm 10mm, clip, width=10 cm]{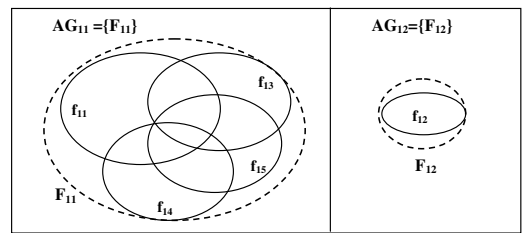}
\caption{\em Example Aggregate Subsets for directive $d_1$}
\label{fig:Fig-AggregateSubset}
\end{figure}
	
\item \textit{Let the core functions, $F_{ij}$, of all the aggregated subsets $AG_{ij}$  related to directive $d_i$ constitute the $i^{th}$ \textbf{Core Function Set}, $CFS_i$, such that $ CFS_i = \{F_{i1}, F_{i2},\ldots,F_{ij}\}; 1 \leq i \leq n; 1 \leq j \leq m $.} Hence, $CFS_i$ comprises core functions that are functional abstractions initially defined at a more detailed level. These functional abstractions are potential Capabilities. Thus, as shown in Figure \ref{fig:Fig-CoreFunctionSet}, $CFS_1=\{F_{11},F_{12}\}$.
\end{enumerate}

Thus, in this manner, the synthesis algorithm defines \textit{a Core Function Set (CFS) for each directive in the system}. Specifically, each directive $d_i$ has an associated $CFS_i$. The elements of a CFS are core functions, which are aggregates derived from a systematic process of synthesizing directives. 

Recall that Capabilities are functional abstractions that exhibit high cohesion and low coupling. Therefore, we now measure the cohesion and coupling values and examine the abstraction level of each core function in order to determine the set of Capabilities.

\begin{itemize}
	\item \textit{Cohesion}:
For each directive the synthesis algorithm generates a CFS comprising core functions. The cohesion of a core function is computed as an average of the relevance values of each participating directive in achieving that function. This implies that the list of directives associated with each core function in every CFS be enumerated; this necessitates substantial time and effort. Also, note that the core functions associated with different directives may be defined at various abstraction levels. Consequently, core functions may be subsets of one another resulting in redundant computations of relevance values. Furthermore, in our empirical analysis we observe that although  the calculation of the average cohesion value is direct, the process of eliciting relevance values for each core function is highly cumbersome and notably subjective. These factors require us to explore alternate approaches for determining the cohesion of potential Capabilities.

\item \textit{Coupling}: 
Units are said to be coupled if changes in a source unit affect one or more dependent entities. The only  information available for computing the coupling between the elements of CFSs in the synthesis algorithm is the set of common directives shared by the core functions. Experimental results show that determining coupling values merely based this number is unrepresentative of the actual implementation. Furthermore, the synthesis approach fails to provide information about the strength of dependency between functions. Hence, we conclude that the synthesis algorithm is ill-equipped to facilitate the computation of coupling between potential Capabilities.

\item \textit{Abstraction Level}: We know that each directive has an associated CFS whose elements are core functions. Empirical analysis reveals that at the abstraction level computed by the synthesis algorithm the core functions of a particular CFS do not share commonalities with other functions. However, any reduction in the abstraction level results in common intersections between aggregates. This is explained by the design of the synthesis algorithm, which terminates once a functional aggregate is established, as illustrated by the example of directive $d_1$. The synthesis algorithm indicates that the abstraction level of a core function is perhaps determined by examining its links with other core functions. 
Therefore, one needs to consider the abstraction level, and the links between aggregates when formulating Capabilities.
\end{itemize}
The synthesis algorithm attempts to identify Capabilities from the detailed directives of complex emergent systems. Given the large magnitude of these systems, considerable effort is required to establish the CFSs for 100s of directives. We note that, although the synthesis algorithm does provide insights regarding an ideal abstraction level of Capability, it is infeasible to automate the computation of cohesion and coupling measures. Therefore, it seems impractical that the synthesis algorithm be utilized for identifying Capabilities. This mandates that we design a more objective algorithm that is far less dependent on user input. Hence, we examine an alternative solution --- a decomposition algorithm based on the top-down approach --- in the following section. 

\section{Decomposition}
\label{sec:Decomposition}

\begin{figure}[htbp]
\centering
\includegraphics[trim = 10mm 173mm 0mm 15mm, clip, width=9.5 cm]{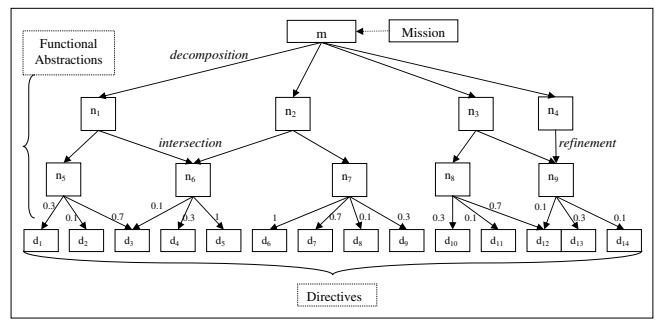}
\caption{\em Example FD Graph $G=(V,E)$}
\label{Fig-Example-Metric}
\end{figure}

The decomposition algorithm utilizes a graph-based representation  of user needs, \textit{viz.} a Function Decomposition (FD) graph, to formulate Capabilities. An FD graph represents functional abstractions of the system obtained by the systematic decomposition of user needs. A need at the highest level of abstraction is the mission of the system and is represented by the root. We use the top-down philosophy to decompose the mission into functions at various levels of abstraction. We claim that a decomposition of needs is equivalent to a decomposition of functions because a need essentially represents some functionality of the system.  Formally, we define an FD graph $G=(V,E)$ as an acyclic directed graph where $V$ is the vertex set and $E$ is the edge set. $V$ represents the system functionality: leaves represent directives, the root symbolizes the mission, and internal nodes indicate system functions at various abstraction levels. Similarly, the edge set $E$ comprises edges that depict decomposition, intersection or refinement relationship between nodes. These edges are illustrated in Figure \ref{Fig-Example-Metric}. An edge between a parent and its child node represents functional decomposition and implies that the functionality of the child is a proper subset of the parent's functionality. Only internal (non-leaf) nodes with an outdegree of at least two can have valid decomposition edges with their children. The refinement relation is used when there is a need to express a node's functionality with more clarity, say, by furnishing additional details. A node with an outdegree of one symbolizes this type of relationship with its child node. To indicate the commonalities between functions defined at the same level of abstraction the intersection edge is used. Hence, a child node with an indegree greater than one represents a functionality common to all its parent nodes. The FD graph utilizes these definitions to provide a structured top-down representation of  system functionality, and thereby, facilitates the decomposition algorithm to formulate Capabilities in terms of their cohesion, coupling, and abstraction values. We discuss the mechanics of the algorithm next.

\subsection{Algorithm}
\label{sec:Slices}
The input to the decomposition algorithm is an FD graph, $G=(V,E)$ that represents the functionality of the system to be developed. We first determine the set of all valid combinations of internal nodes that can be considered as potential Capabilities. These combinations are termed $slices$. Then we compute the cohesion and coupling measures for each slice and examine the levels of abstraction to establish the finalized set of Capabilities. 

We define slice $S$ as a subset of $V$ where the following constraints are satisfied:

\begin{enumerate}
	\item \textit{Complete Coverage of Directives:} We know that a Capability is associated with a set of directives, which is finally mapped to system requirement specifications (see Figure \ref{fig:FigNeedsToRequirements}). Consequently, a set of Capabilities of the system has to encompass all the directives derived from user needs. The leaves of the FD graph constitute the set of all directives in a system. We ensure that each directive is accounted for by some Capability, by enforcing the constraint of complete coverage given by 
${\overset{m}{\underset{\scriptstyle i=1} \bigcup} D_i}=L$, where
\begin{itemize}
	\item $D_i$ denotes the set of leaves associated with the $i^{th}$ node of slice $S$
	\item	$L=\{u \in V| outdegree(u)=0\}$ denotes the set of all leaves of $G$
	\item	$m=|S|$ 
\end{itemize}

\item \textit{Unique Membership for Directives:} In the context of directives, by ensuring that each directive is uniquely associated with exactly one Capability, we avoid implementing redundant functionality. Otherwise, the purpose of using slices to determine Capabilities as unique functional abstractions is defeated. We ensure the unique membership of directives by the constraint ${\overset{m}{\underset{\scriptstyle i=1} \bigcap} D_i}=\{\phi\}$. 

\item \textit{System Mission is not a Capability:} The root is the high level mission of the system and cannot be considered as a Capability. The  cardinality of a slice containing the root can only be one. This is because, including other nodes with the root in the same slice violates the second constraint. 
Hence, $\forall u \in S, indegree(u) \neq 0$. 

\item \textit{Directive is not a Capability:} A leaf represents a directive, which is a system characteristic. A slice that includes a leaf fails to define the system in terms of its functionality and focuses on describing low level details. 
Hence, $\forall u \in S, outdegree(u) \neq 0$.
\end{enumerate}

The cohesion and coupling values for each slice is computed using the measures described next. We also discuss the average abstraction level of nodes that possess high cohesion and low coupling values.

\begin{itemize}

\item \textit{Cohesion}:
As in the synthesis algorithm, the cohesion of a node in a slice is computed as an average of the relevance values of the participating directive. The relevance values are assigned based on the values listed in Table \ref{tab:ScaleOfRelevance}. However, we make a distinction between the parent and ancestor nodes of a directive. In order to reduce the need for user input, we elicit the relevance value of a directive only with respect to its parent node, whose cohesion is the arithmetic mean of the relevance values of its directives. Figure \ref{Fig-Example-Metric} illustrates relevance values of directives to their parents. However, the cohesion of an ancestor is computed as a weighted average of the size (number of associated directives) and cohesion of its non-leaf children. Specifically, the cohesion measure of an internal node $n$ with $t>1$ non-leaf children is:
\begin{displaymath}
Ch(n) =	 \dfrac{\overset{t}{\underset {\scriptstyle i=1} \sum} (size(v_i). 					Ch(v_i))}{\overset{t} {\underset {\scriptstyle i=1} \sum} size(v_i)}	
\end{displaymath}

such that $(n,v_i)\in E $  and,
					\begin{displaymath}
				size(n)= \begin{cases}
									\overset{t}{\underset {\scriptstyle i=1} \sum } size(v_i) & (n,v_i) \in E; outdegree(v_i)>0;\\
								 1 & outdegree(n)=0 
									\end{cases}
					\end{displaymath}
\item \textit{Coupling:}
To measure coupling we need information about dependencies between system functionalities. By the virtue of its construction, the structure of the FD graph represents the relations between different aggregates. In particular, we compute coupling between two nodes in a slice in terms of their directives. Two directives are said to be coupled if a change in one affects the other. We compute this effect as the probability that such a change occurs and propagates the shortest path ($dist$) between them. Note that the coupling measure is asymmetric. Generalizing, the coupling measure between any two internal nodes $p, q \in V$, where $outdegree(p)>1, outdegree(q)>1$ and $D_p \cap D_q=\{\phi\}$ is:
\begin{center}
$Cp(p,q)= \dfrac{\displaystyle \sum_{d_i \in D_{p}} 
\displaystyle  \sum_{d_j \in D_{q}} Cp(d_i,d_j)} {|D_{p}|. |D_{q}|} $
\end{center}

where $Cp(d_i,d_j)= \dfrac{P(d_j)}{dist(d_i,d_j)}$ and $P(d_j)= \dfrac{1}{|D_q|}$. \\

$P(d_j)$ is the probability that directive $d_j$ changes among all other directives associated with the node $q$. 
 
\item \textit{Abstraction Level:}
The experimental results of the decomposition algorithm indicate that slices with nodes that exhibit maximum cohesion and decreased coupling are also at higher abstraction levels. We know that abstraction level is related to size; the higher the level of a node, the greater the number of its associated directives. Thus, the decomposition algorithm identifies Capabilities as nodes that exhibit high cohesion and low coupling but are also of increased sizes, which is undesirable from an implementation standpoint.
\end{itemize}

The decomposition algorithm provides an approach to automate the cohesion and coupling measures. Preliminary experimental results indicate that values computed using these metrics are indicative of desirable software engineering characteristics. In particular, we observe that  on an average, in a slice, nodes \textit{viz.} Capabilities that have high cohesion values also exhibit low coupling
with other nodes. However, the decomposition approach fails to provide nodes at an abstraction level that are optimal with respect to size. Therefore, we now explore a reconciliation between the synthesis and decomposition algorithms to determine Capabilities that are optimal with respect to the abstraction levels and the computations of cohesion and coupling.

\section{Reconciliation}
\label{sec:Reconciliation}

Sections \ref{sec:Synthesis} and \ref{sec:Decomposition} describe the synthesis and decomposition algorithms to formulate Capabilities.  In particular, we observe that the computation of coupling and cohesion values using the \textit{decomposition approach} can be easily automated. This is because the coupling measure is a function of distance of change propagation and probability of change, and therefore, is completely objective. Likewise, the cohesion measure, although less objective, is conveniently computed for all functional abstractions. In contrast, the excessive subjectivity of the synthesis approach presents little scope for automating the formulation of Capabilities in complex emergent systems. However, unlike the decomposition algorithm, the \textit{synthesis approach} provides insights about the optimum abstraction level of a Capability. Hence, we construct a composite algorithm to formulate Capabilities such that it incorporates elements of cohesion and coupling from the decomposition algorithm and that of the abstraction level from the synthesis algorithm.  In this section, we first enumerate the steps of the composite algorithm. Then we use the example of the computer based library system to illustrate the reconciliation of the top-down and bottom-up approaches in the composite algorithm.

\subsection{Composite Algorithm}
The composite algorithm represents system functionalities at different abstraction levels using an FD graph, as in the decomposition approach. 
This is because, in the synthesis approach one needs to consider all possible system directives and then determine functional abstractions through an iterative process, which challenges the limited processing capacity of the human mind \cite{Miller1956}. In contrast, the decomposition algorithm provides a more structured approach and begins with a single entity --- system mission --- that is easily comprehensible. Therefore, the input to the composite algorithm is an FD graph representation of the system functionality.  The steps of the algorithm are detailed below:
\begin{enumerate}
	\item Construct an FD graph to represent the system functionality. 

\begin{figure*}[htbp]
\centering


\includegraphics[trim = 10mm 0mm 0mm 10mm bb=25 23 587 307 ]{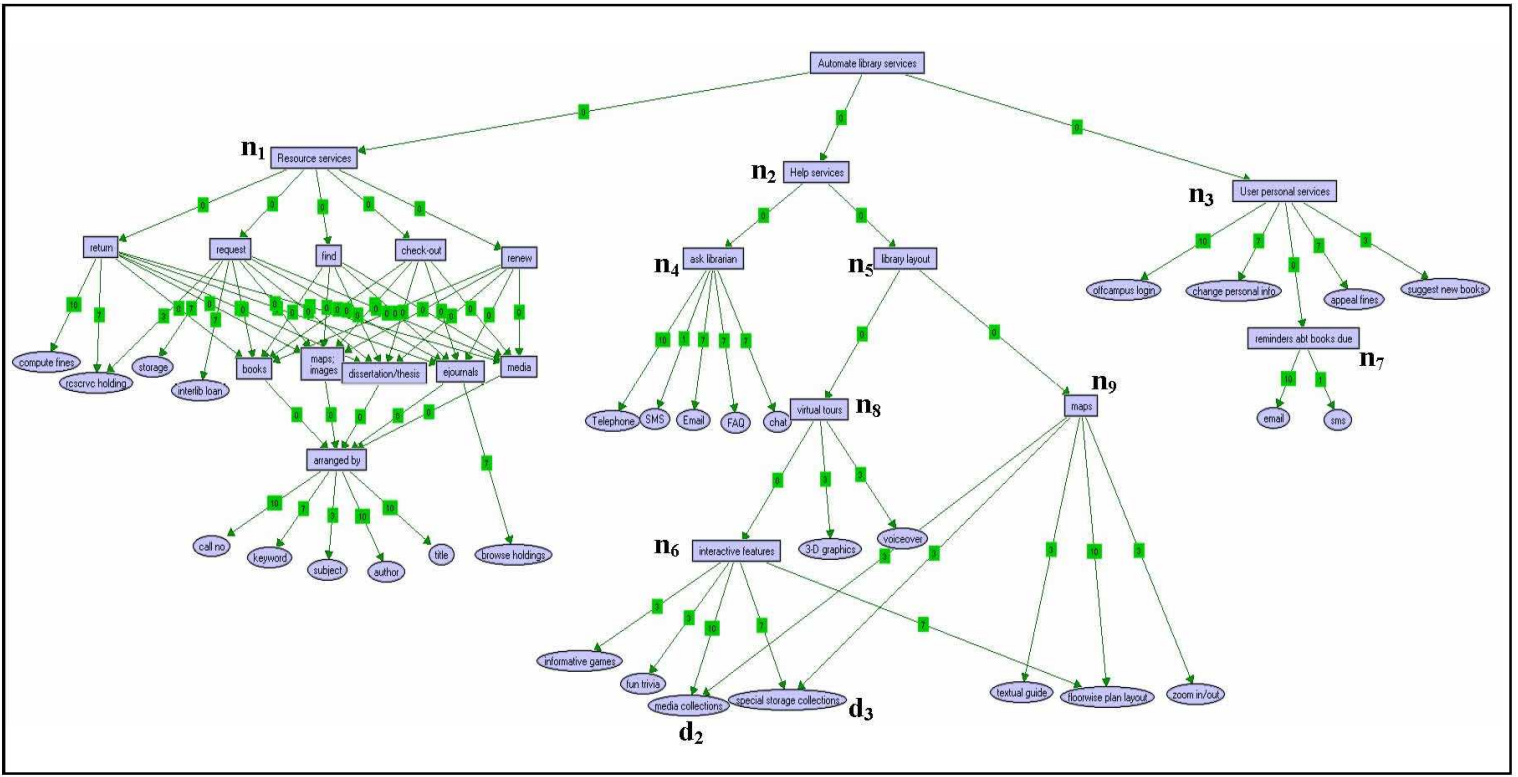}

\caption{\em FD Graph of Library System}
\label{fig:Fig-library}
\end{figure*}
	
	\item Determine all possible slices from the FD graph. 
	Each node within a slice is associated with a unique set of directives such that the union of these directive sets is equivalent to entire set of directives of a system. 
	\item Compute the cohesion and coupling values for nodes in each slice using the metrics defined by the decomposition approach, described in Section \ref{sec:Decomposition}.
	 Use these values to determine the average cohesion and coupling measures of a slice. 
	\item Similarly, compute the abstraction levels and sizes of each internal node in a slice.
\item The set of Capabilities is that slice which exhibits high cohesion, low coupling and comprises nodes of balanced abstraction levels. 
\end{enumerate}

We now illustrate the steps of the composite algorithm using the example of the library system.

\subsubsection{Constructing FD Graph}
In accordance with the first step of the composite algorithm an FD graph is constructed for the library system. To begin the process of systematic decomposition, we first identify the overall mission: develop software to automate library services. The mission is represented by the root node of the FD graph, which is illustrated in Figure \ref{fig:Fig-library}. Observe that the mission is partitioned into functionalities of lower abstraction levels. The elliptical vertices in the FD graph denote the directives of the library system and the internal nodes indicate potential Capabilities. Also, the weight associated with an edge symbolizes the relevance value of a directive to its immediate parent node. Figure \ref{fig:Fig-library} shows these values as $1,3,7,$ or $10$. However, they are normalized on a [0,1]  scale (as defined in Table \ref{tab:ScaleOfRelevance})  for the computation of cohesion measure. Note that the weight of an edge between internal nodes is inconsequential and so is denoted by zero.
 
 \subsubsection{Determining Slices}
In our experiment with the library system we computed $1014$ valid slices from a possible $1048576$ combinations of nodes. In essence there are six basic sets of slices and these are listed in Table \ref{tab:SliceSet}. Permutations of each basic slice set are also valid combinations. For example, for the slice $S_1$ of Table \ref{tab:SliceSet}, the permutations $\{n_1,n_2,n_3\}$ and $\{n_1,n_3,n_2\}$ are also considered as unique slices. This is primarily because the coupling measure is asymmetric, \textit{i.e.} $Cp(n_i,n_j) \neq Cp(n_j,n_i)$, $i \neq j$; coupling is a function of probability of change which is computed using the size of a node. Consequently, permutations of the basic slice sets also need to be considered as individual slices. We conjecture that the coupling measure can assist in choosing an implementation order of Capabilities that potentially minimizes the impact of change.
\begin{table}[htbp]
\centering
\caption{\em Basic Slice Sets}
\begin{tabular}{|l|c|c|}
\hline
\textsc{Slice} & \textsc{Nodes} & \textsc{Permutations}\\
\hline 
\hline
$S_1$ & ${n_1,n_2,n_3}$ & $2^3$ \\
\hline
$S_2$ & ${n_1, n_3, n_4,n_5}$ & $2^4$ \\
\hline
$S_3$ & ${n_1, n_3, n_4,n_8,n_9}$ & $2^5$ \\
\hline
$S_4$ & ${n_2,n_3,n_{10},n_{11}}$ & $2^4$ \\
\hline
$S_5$ & ${n_3,n_4,n_5,n_{10},n_{11}}$ & $2^5$ \\
\hline
$S_6$ & ${n_3, n_4,n_8, n_9,n_{10},n_{11}}$ & $2^6$ \\
\hline

\end{tabular}
\label{tab:SliceSet}
\end{table}

An interesting observation from the FD graph of the library system is that nodes $n_6$ and $n_7$ are the only nodes that are not a part of any slice. Further analysis reveals the following explanation: Nodes $n_6$ and $n_7$ are internal nodes with only directives as their siblings. Moreover, the parents of $n_6$ and $n_7$ are internal nodes whose children are directives and internal nodes. The constraint of a slice definition, \textit{viz.} unique membership of directives, disallows $n_6$ and $n_8$ or $n_7$ and $n_3$ from being a part of the same slice. In addition, the requirement for complete coverage of directives of nodes in a slice necessitates the exclusion of $n_6$ and $n_7$ from any slice. However, it is possible that $n_6$ or $n_7$ can be the root of a large subgraph, in which case their exclusion is detrimental to the formation of Capabilities. This observation compels one to explore the relationship and distribution of internal nodes and directives, when defined at the same level.

\subsubsection{Computations}
Once the slices are determined we compute the cohesion, coupling and abstraction values for each node of a slice. The arithmetic mean of these values provides statistics that help ascertain the quality of a slice from a software engineering perspective. Before discussing the computations, however, we first analyze the relationship between a node's abstraction level, its depth and its size. This is because the objective of the composite algorithm is to formulate Capabilities that not only exhibit high cohesion and low coupling, but are also of balanced abstraction levels. Furthermore, unlike abstraction level, cohesion and coupling are well established concepts that are accepted by the software engineering community. 
Therefore, it is imperative that we present our notion of an ``abstraction level'' using insights provided by the synthesis approach and discuss its role in identifying Capabilities.
\begin{itemize}
	\item \textit{Abstraction Level, Depth and Size:}
An abstraction presents information essential to a particular purpose by omitting irrelevant elements. A Capability is a functional aggregate that indicates the functionality expected of the system from a high-level perspective while ignoring minute details. Similarly, the mission of a system is a functional aggregate described at the highest level of abstraction because it states the overall system functionality \textit{sans} low-level information. Therefore, with respect to the FD graph, the root is of the highest abstraction and a directive is the lowest. Furthermore, we observe that the abstraction of the internal nodes decreases with their increased distance from the root; distance is the length of the shortest path from the root. We refer to this distance as the \textit{depth} of a node. More specifically, the greater a node's depth, the lower is its abstraction level. Therefore, for a node $n \in V$ in an FD Graph $G$, the qualitative relation between abstraction level and depth is denoted as:
\begin{displaymath}
Abstraction(n) \propto \dfrac{1}{depth(n)}
\end{displaymath}
We now discuss the relation between a node's depth and its size. Recall that size is the number of  directives associated with a node. The FD graph of the library system indicates that the size of an internal node decreases as its depth increases. For example, in Figure \ref{fig:Fig-library} node $n_2$ has size $14$ and depth $1$ whereas $n_4$ is of size $5$ and depth $2$. We confirm if the random variables, size and depth, are truly correlated by using a scatter plot. Specifically, we plot the average values of size and depth of a node within a slice. This data is obtained from the 1014 slices computed from the FD graph of the library system. 
The scatter plot is shown in Figure \ref{fig:Fig-ScatterPlot} where, within a given slice, the average size of a node is plotted against the average depth of a node.
\begin{figure}[htbp]
\centering
\includegraphics[trim = 22mm 180mm 20mm 20mm, clip, width=11 cm]{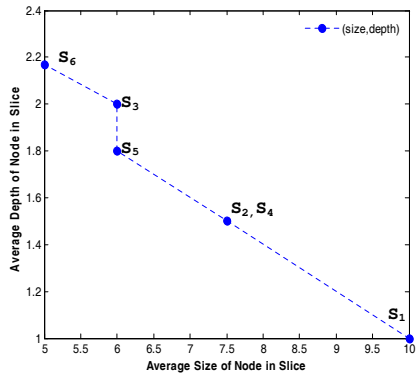}
\caption{\em Illustrating the relation between depth and size}
\label{fig:Fig-ScatterPlot}
\end{figure}
We discuss the following observations about the scatter plot diagram:
\paragraph{Permutations}
 In the scatter plot of Figure \ref{fig:Fig-ScatterPlot}, each data point with coordinates $(size,depth)$ corresponds to the basic slice set listed in Table \ref{tab:SliceSet}, and therefore, is denoted by $S_i$.  Although we plot the average size and the depth of a node in each of the 1014 slices computed from the FD graph of the library system, there are only five data points. This is because the permutations of the basic slices results in additional valid slices. However, the average node size and depth values remain unchanged for the permutations of the same basic slice set. Therefore, each point $S_i, (i=1, \ldots, 6)$ represents the average $(size,depth)$ of a node, which is unchanging in the permutations corresponding to the basic slice sets, shown in Table \ref{tab:SliceSet}.
	
\paragraph{Correlation}
The scatter plot shows that there is a relation between the depth and size, which is reaffirmed by the large value of their correlation coefficient $Corr(size,depth)=-0.966630075$. This implies that depth and size are highly negatively correlated. In addition, as discussed earlier, we know that the level of abstraction decreases with an increase in  depth. Using this relation and the negative correlation between size and depth, one infers that the size of a node is proportional to its abstraction level. Therefore, for a node $n \in V$ in an FD graph $G=(V,E)$, the qualitative relation between abstraction level and size is denoted as:
\begin{displaymath}
Abstraction(n) \propto size(n)
\end{displaymath}
Hence, the relation between a node's size, depth and abstraction level is used to assist in judicious identification of  Capabilities. In particular, we conclude that a balanced abstraction level is influenced by the size of a node.

\item \textit{Cohesion and Coupling:}
The cohesion and coupling measures are computed using the metrics defined in Section \ref{sec:Decomposition}. The maximum, minimum, and median average cohesion and coupling values of the slices in the library system are also detailed in Table \ref{tab:ChCp}, along with the average size and depth values. 
\begin{table}[htbp]
\centering
\caption{\em Average Values of Slices of Library System}
\begin{tabular}{|l|c|c|c|c|}
\hline
\textsc{Range} & \textsc{Cohesion} & \textsc{Coupling} & \textsc{Size} & \textsc{Depth}\\
\hline \hline
Maximum & $0.65119$ & $3.58603$ & $10$ & $2.16667$\\
\hline
Minimum & $0.599048$ & $0.547891$ & $5$ & $1$\\
\hline
Median & $0.599048$ & $2.76445$ & $5$ & $2.16667$\\
\hline
\end{tabular}
\label{tab:ChCp}
\end{table}
\end{itemize}
\subsubsection{Selecting Capabilities}
\label{sec:SelectingCapabilities}
We illustrate the final step of the composite algorithm --- determining the optimum set of Capabilities from the set of all slices --- using the example library system. Of specific interest are slices that exhibit, on an average, high cohesion and low coupling values. In particular, among all slices computed from the FD graph (Figure \ref{fig:Fig-library}) of the library system, we examine the slices, $PS_1$ and $PS_2$, that have the two highest cohesion values. $PS_1$ and $PS_2$ are permutations of the basic slice sets $S_1$ and $S_2$ respectively. 
The cohesion and coupling values of slices $PS_1$ and $PS_2$  are presented in Table \ref{tab:Slices}. Note that the cohesion and coupling values of $PS_1$ and $PS_2$ are higher and lower respectively, than the median values of the slices of the library system, described in Table \ref{tab:ChCp}. In particular, slice $PS_1$ has maximum cohesion and minimum coupling values. Slice $PS_2$ exhibits the second highest cohesion value and a coupling measure of $1.09577$ that is lower than the overall coupling median of $2.76445$.

\begin{table}[h]
\centering
\caption{\em Average values of Slices $S_1$, $S_2$ and $S_3$}
\begin{tabular}{|l|c|c|c|}
\hline
\textsc{Slice} & \textsc{Cohesion} & \textsc{Coupling} & \textsc{Nodes} \\
\hline \hline
$PS_1$ & $0.65119$ & $0.547891$ & $\{n_3, n_1, n_2\}$ \\
\hline
$PS_2$ & $0.636873$ & $1.09577$ & $\{n_4, n_3, n_1, n_5\}$ \\
\hline
$PS_3$ & $0.603689$ & $1.59961$ & $\{n_4, n_9, n_3, n_8, n_1\}$ \\
\hline
\end{tabular}
\label{tab:Slices}
\end{table}
According to the decomposition algorithm, slice $PS_1$ is the most optimal among all slices of the library system. This is because it exhibits maximum cohesion and minimum coupling values when compared to all other slices. In contrast, the synthesis algorithm chooses slice $PS_2$ as the desirable set of Capabilities. Recall that the synthesis approach emphasizes on low abstraction levels, and consequently, reduced sizes.
The average size of each node in $PS_1$ is $10$ where as that of $PS_2$ is $7.5$ as shown in Table \ref{tab:SlicesSizeLevel}. The implementation size of individual aggregates in $PS_1$ can be reduced if node $n_2$ is replaced by $n_4$ and $n_5$, which results in a combination of nodes $(n_1, n_4, n_5, n_3)$, \textit{viz.} basic slice set $S_2$ and in particular, the least coupled permutation, $PS_2$.	This also implies that nodes $n_4$ and $n_5$ are at a lower level of abstraction than node $n_2$. The marginal increase in coupling of $PS_2$ is offset by the advantage of constructing smaller sized Capabilities. We observe from the FD graph in Figure \ref{fig:Fig-library} that there are no intersection edges between nodes $n_4$ and $n_5$, signifying that this is a balanced abstraction level. In contrast, let us consider the scenario where we choose to implement nodes, say $n_8$ and $n_9$ instead of $n_5$, which are defined at much lower level of abstraction and are of a smaller size. The node combination $(n_1, n_4, n_8, n_9, n_3)$ is actually the basic slice set $S_3$ listed in Table \ref{tab:SliceSet}. We choose the permutation $PS_3$=$\{n_4, n_9, n_3, n_8, n_1\}$ because it has the lowest coupling among all possible permutations of $S_3$. Recall that all permutations exhibit the same cohesion, and hence, fails to influence the selection. The cohesion, coupling, size and depth values values are shown in comparison with $PS_1$ and $PS_2$ in Table \ref{tab:Slices} and Table \ref{tab:SlicesSizeLevel}.
\begin{table}[h]
\centering
\caption{\em Average Size and Depth}
\begin{tabular}{|l|c|c|}
\hline
\textsc{Slice} & \textsc{Average Size} &  \textsc{Average Depth}  \\
\hline \hline
$PS_1$ &  $10$ & $1$ \\
\hline
$PS_2$ & $7.5$ & $1.5$ \\
\hline
$PS_3$ & $6$ & $2$  \\
\hline
\end{tabular}
\label{tab:SlicesSizeLevel}
\end{table}
Slice $PS_3$ has a lower cohesion and a higher coupling value than $PS_1$ and  $PS_2$. However, it exhibits values that is better than the median cohesion and coupling values of the overall slices of the library system. In addition, $PS_3$ also has a smaller size when compared to the selections of the decomposition algorithm --- $PS_1$--- or the synthesis algorithm --- $PS_2$. Collectively, these factors seem to indicate that this slice is perhaps a more optimal choice than either $PS_1$ or $PS_2$ --- as illustrated by Table \ref{tab:SlicesSizeLevel}. However, we observe from the FD graph in Figure \ref{fig:Fig-library} that nodes $n_8$ and $n_9$ share common directives $d_2$ and $d_3$. If these nodes are considered as Capabilities, then directives $d_2$ and $d_3$ will be associated with  only one Capability. This implies that linkages will be broken to ensure that each directive is bound to either $n_8$ or $n_9$. Even so, this does not eliminate the implicit coupling that exists between $n_8$ and $n_9$. Consequently, it is difficult to consider nodes $n_8$ and $n_9$ as Capabilities with reduced coupling. This nonconformity with a basic tenet of the Capability definition forces us to reject $PS_3$ as an optimal set, and instead,  choose slice $PS_2$. 

To summarize, the composite algorithm constructs an FD graph, determines all possible slices and utilizes the cohesion and coupling computations of the decomposition algorithm to produce potential sets of Capabilities. Then, it incorporates the criteria of a balanced abstraction level and reduced sizes as illustrated by the synthesis algorithm to choose an optimal slice for the library system.
Thus, the elements of cohesion and coupling from the top-down approach and the abstraction level from the bottom-up approach constitute the composite algorithm to determine Capability sets for developing change-tolerant systems.

\section{Conclusion}
\label{sec:Conclusion}
Software engineering methods for system development are often based on a top-down or a bottom-up approach. However, our solution framework, CE, constructs complex emergent systems as change-tolerant entities by utilizing a complementary approach. In particular, we use a composite algorithm to formulate Capabilities as maximally cohesive and minimally coupled functional abstractions of a system. Presently, a Capability depicts only the functionality of a system; integrating non-functional aspects into the definition of a Capability is part of future work. The cohesion and coupling measures of these basic building blocks are computed as in the decomposition algorithm and the abstraction level as defined by the synthesis algorithm. Note that the former algorithm is based on a top-down approach while the latter, on a bottom-up approach. Thus, the composite algorithm is a blend of the two polar approaches. Experimental results further substantiate the need for such a complementary approach. Our experience in assessing the efficacy of the synthesis and decomposition algorithms aids in understanding the essence of a Capability, and emphasizes that the design of Capabilities is in fact a reconciliation of diametrically opposite approaches to problem solving. 

\bibliography{ECBS07}
\end{document}